\documentclass[a4paper,11pt]{article}
\usepackage[dvipdfmx]{graphicx}
\usepackage{comment}
\usepackage{jinstpub} 

\title{\boldmath Calibrating the photometric performance of a high-time-resolution photon-counting imager for optical astronomy}

\author[a]{Mana~Hasebe,}
\author[a]{Takeshi~Nakamori,}
\author[b,a]{Kazuaki~Hashiyama,}
\author[a]{Anju~Sato,}
\author[a]{Miu~Maeshiro,}
\author[a]{Rin~Sato,}
\author[c]{Masayoshi~Shoji,}
\author[d]{Masaru~Kino,}
\author[e,f]{Dai~Takei,}
\author[a]{Tomohiro~Sato,}
\author[g]{Kazuki~Ueno}

\affiliation[a]{Yamagata University,\\
1-4-12 Kojirakawa, Yamagata 990-8560, Japan}
\affiliation[b]{The University of Tokyo,\\
7-3-1 Hongo, Bunkyo, Tokyo 113-8654, Japan}
\affiliation[c]{KEK,\\
1-1 Oho, Tsukuba, Ibaraki 305-0801, Japan}
\affiliation[d]{Kyoto University, Okayama Observatory,\\
3037-5 Honjo, Kamogata, Asakuchi, Okayama 719-0232, Japan}
\affiliation[e]{Daiphys Technologies LLC,\\
1-5-6 Kudan-Minami, Chiyoda, Tokyo 102-0074, Japan}
\affiliation[f]{Rikkyo University,\\
3-34-1 Nishi-Ikebukuro, Toshima, Tokyo 171-8501, Japan}
\affiliation[g]{Osaka University,\\
1-1 Machikaneyama, Toyonaka, Osaka 560-0043, Japan}

\emailAdd{s242044m@st.yamagata-u.ac.jp, nakamori@sci.kj.yamagata-u.ac.jp}

\abstract{
Optical observations with high time resolution are essential for understanding the origin of sub-millisecond timescale astronomical phenomena,
including giant radio pulses from the Crab Pulsar. 
We have developed a high-speed imaging system called the Imager of MPPC-based Optical photoN counter from Yamagata (IMONY).
The system uses a customized Multi-Pixel Photon Counter (MPPC), which independently reads out signals from all $8\times8$~pixels and functions as an imager based on a Geiger-mode avalanche photodiode array.
This system assigns timestamps to detected photons with a time resolution of $100\,\mathrm{ns}$.
We installed IMONY on the $3.8\,\mathrm{m}$ aperture Seimei Telescope in Okayama, Japan. 
We have successfully detected the 34-ms period of the Crab Pulsar 
and imaged stars in the sensor's field of view. 
However, we have also found that a small fraction of the pixels have shown double or multiple pulses that are used for photon arrival timing.
This situation is likely due to circuit noise and may unfortunately result in overestimating the number of photons detected.
In order to precisely estimate the photon flux of targets or the sky background, 
calibration of such over-counts is important. 
We measured the number of detected photons relative to the light intensity of each pixel in a laboratory environment. 
We estimated the number of spurious hit pulses caused by signal tail fluctuations exceeding the comparator threshold, based on the exponential distribution of time intervals between pulses. These are distinct from typical SiPM afterpulses and originate from electronic effects in our readout system.
After applying the calibration to the observed data, 
we confirmed the linearity between the V-band magnitudes of stars and the number of detected photons.
}


\begin{document}
\maketitle
\flushbottom

\section{Introduction}
\label{sec:intro}


Simultaneous observations across multiple wavelengths are essential for elucidating the radiation mechanisms of non-thermal phenomena with rapid variations in astrophysical objects.
The Crab Pulsar, a rapidly rotating neutron star, 
is known for its pulsed emission spanning from radio to gamma rays. 
In addition, Giant Radio Pulses (GRPs), characterized by extreme rise of intensity in ns-$\mu$s time scales, 
are observed enhanced X-ray and optical emissions associated with GRPs \cite{ref:enoto}.
Studying the detailed time variability of this phenomenon at multi-wavelengths is expected to constrain radiation models.
In the optical band, Complementary Metal-Oxide Semiconductor (CMOS) cameras 
are widely used for high-cadence observations,
although the time resolution is not enough for the GRP studies.
For instance, the CMOS camera on the Kiso 1.0-m Schmidt Telescope 
achieves the time resolution at most $\sim 2$\,ms\cite{ref:cmos} even in a partial readout mode.
Previous research and developments in photon-counting techniques (e.g., \cite{ref:maz, ref:kan, ref:zam})
have shown promise in achieving sub-millisecond time resolution in the optical band.
Although sub-microsecond timing with single-photon detectors has been demonstrated in previous studies (see references \cite{ref:nakamori, ref:kanata}), their application in photon-counting imaging systems for optical astronomy remains limited in practice. While CMOS-based cameras are limited by their readout electronics to a few milliseconds, photon-counting systems allow timestamping at nanosecond resolution. However, in actual observations, the achievable time resolution is ultimately constrained by photon statistics and observing conditions. We aim to facilitate time-domain optical astronomy on short timescales by introducing a compact photon-counting imaging system.

\begin{figure}[htbp]
\centering
\includegraphics[width=0.95\textwidth, trim={35 285 35 285}, clip]{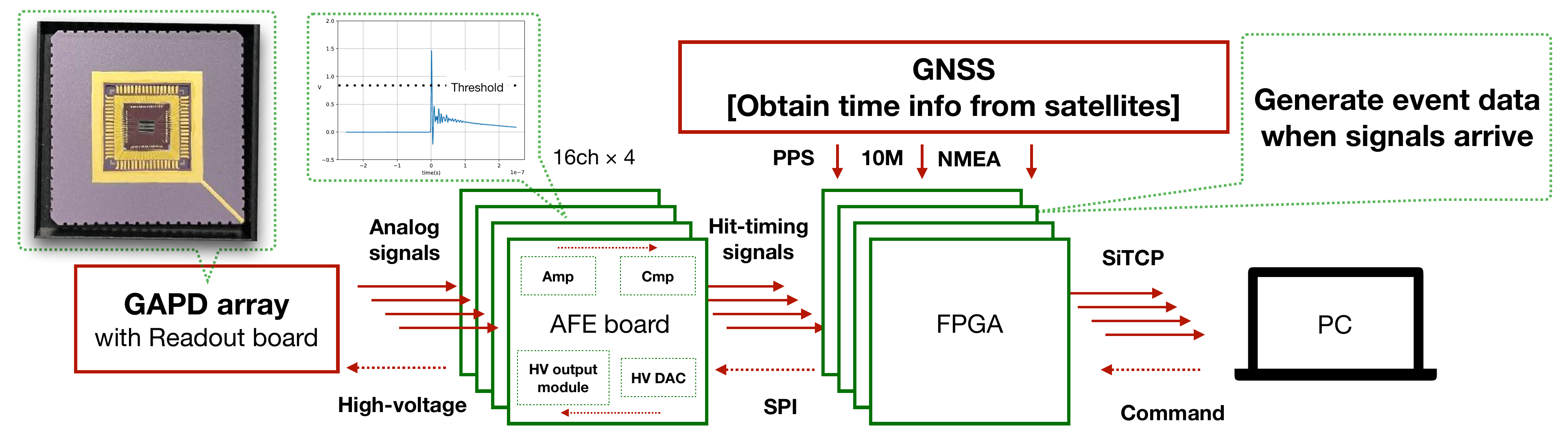}
\caption{The overview of the IMONY system and the signal processing flow. }
\label{fig:imosys}
\end{figure}

Given this context,
we have developed the Imager of MPPC-based Optical photoN counter from Yamagata (IMONY), 
a high-speed imaging system that utilizes a Geiger-mode avalanche photodiode (GAPD) array as a sensor.
This system employs an $8\times 8$ GAPD array 
that is manufactured by Hamamatsu Photonics K.K. and is sensitive enough to detect single photons. 
The maximum photon detection efficiency (PDE) is $\sim70\%$ at 480~nm.
In this study, we primarily used the sensor with the pixel size of 200~$\mu$m.
IMONY works as an imager since the system can read out each pixel individually.
The readout system originally developed for a 16-pixel sensor \cite{ref:nakamori, ref:kanata} 
has been extended to read out 64 pixels sensor \cite{ref:hashi}.
Figure~\ref{fig:imosys} shows an overview of the system, including its signal processing flow.
This system consists of the sensor, analog frontend (AFE) boards,
a GNSS receiver (FURUNO~GF-8803), and commercial FPGA evaluation boards (DIGILENT NEXYS~A7). 
When the GAPD array detects a photon, 
the corresponding analog pulse signal is converted into hit timing signals at the AFE board. 
A timestamp is then assigned to the hit signal by the FPGA with a time resolution of $100\,\mathrm{ns}$ to generate event data.
By combining this information with the GNSS-derived time reference,
an absolute timing of the pulse detection is reconstructed.
The data is transmitted to a PC via Ethernet using SiTCP \cite{ref:sitcp}.

We installed IMONY on the 3.8-m Seimei telescope,
one of the largest aperture telescopes in East Asia, located in Okayama, Japan.
Using this system, we successfully imaged stars within the sensor's field of view 
and detected periodic optical emission from the Crab Pulsar \cite{ref:hashi, ref:hashiD}. 
The Crab Pulsar is a neutron star that rotates with a period of 34~ms, 
producing pulsed emission observed across multiple wavelengths.
The precise period detection by IMONY demonstrates the capability to study GRPs,
while highlighting the importance of evaluating photometric accuracy.

\section{Correction for spurious hit pulses}


\begin{figure}[b]
\centering
\begin{minipage}[b]{0.45\columnwidth}
    \centering
    \includegraphics[width=0.9\columnwidth]{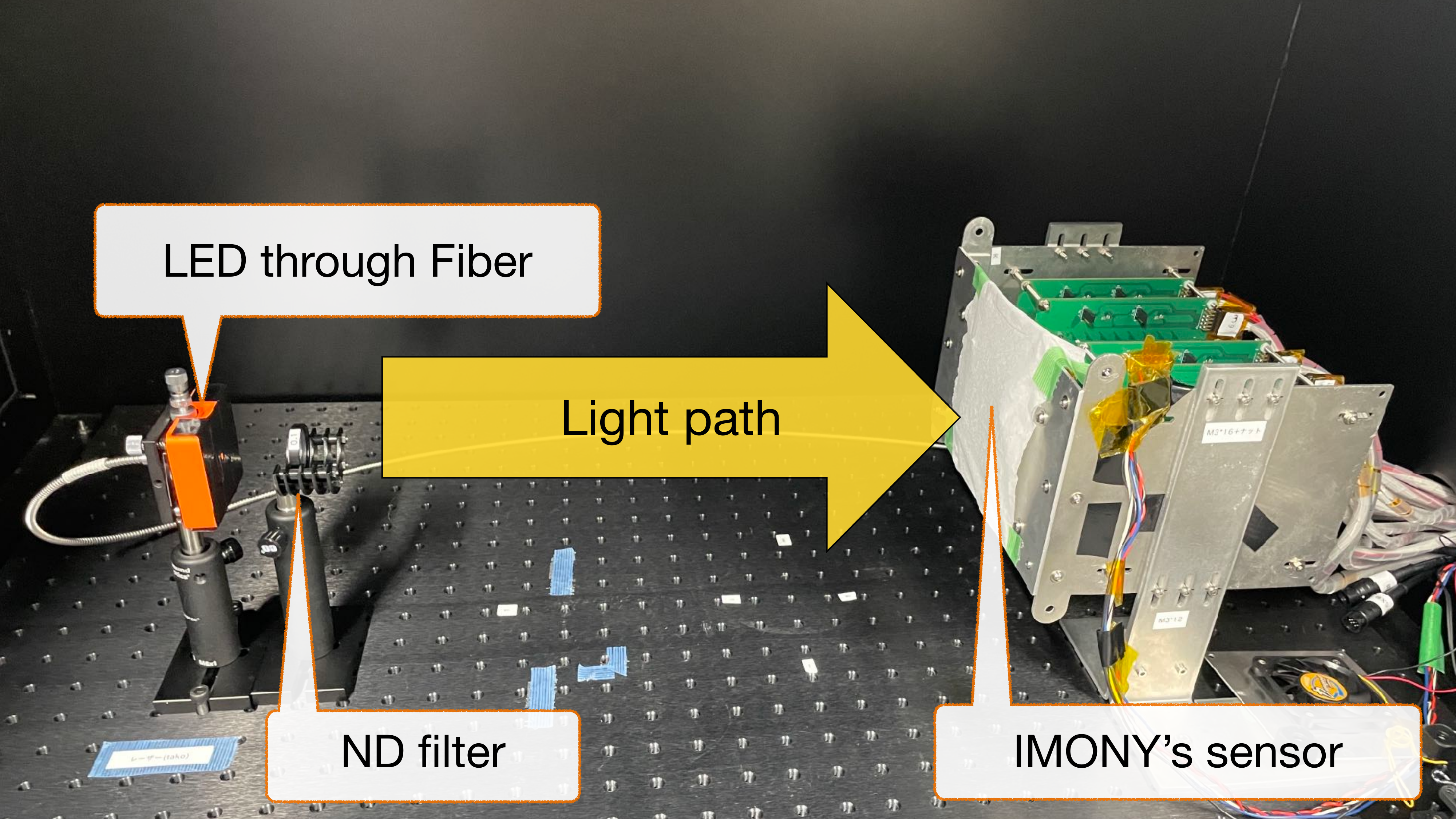}
    \caption{Setup photo of the experiment to measure the linearity of the sensor response. 
    The DC light intensity was adjusted using neutral density filters,
    and sensor was covered to ensure uniform illumination.}
    \label{fig:exsetup}
\end{minipage}
\hspace{0.01\columnwidth}
\begin{minipage}[b]{0.45\columnwidth}
    \centering
    \includegraphics[width=0.9\columnwidth, trim={325 55 325 65}, clip]{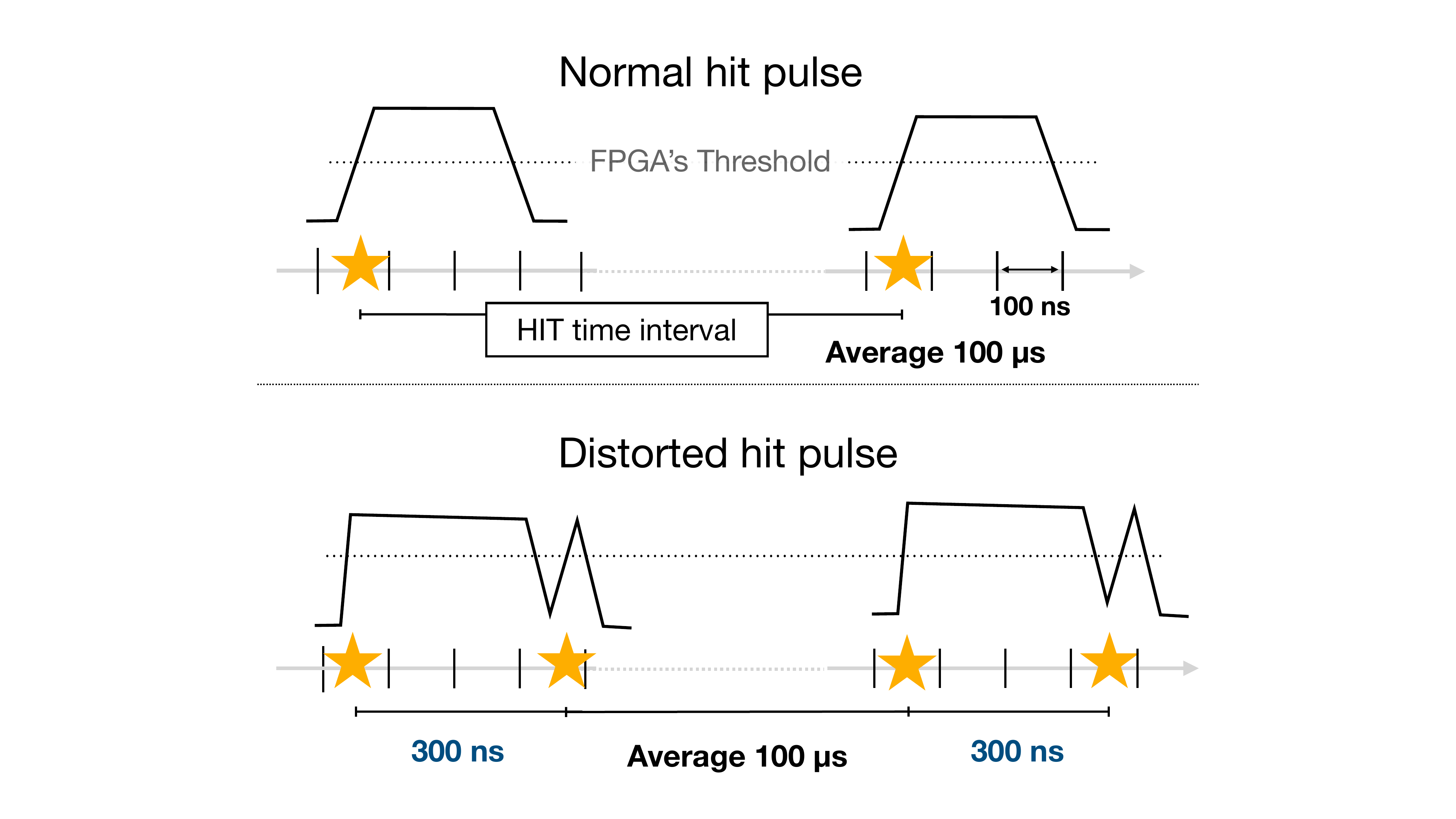}
    \caption{Schematic diagram showing the time interval between normal and distorted hit pulses.}
    \label{fig:hakei}
\end{minipage}
\end{figure}

In optical observations, the number of arriving photons may fluctuate  over time due to changing atmospheric conditions, making flux estimates dependent on comparisons with reference stars captured with the observation target. 
To enhance flux estimation accuracy, we evaluated the detector's linear response to incident light under controlled laboratory conditions.
Figure~\ref{fig:exsetup} shows the experimental setup inside a dark box.
The operating temperature was room temperature and was monitored using a BME280 sensor (Bosch Sensortec).
During verification, we identified an issue: 
A few pixels exhibited double or multiple pulses in the hit-timing signals shown in Figure~\ref{fig:hakei}, likely due to an excessively low comparator threshold setting.
Although contributions from afterpulses at $>300$\,ns timescales cannot be ruled out entirely,
the observed spurious pulses are primarily attributed to signal tail fluctuations crossing the threshold.
Such pulses could lead to an overestimation of photon flux.
To address this issue, we considered two correction strategies,
both incorporating flat field correction (FFC), 
a standard technique in optical astronomy to compensate for pixel-to-pixel variations.
We acquired imaging data under uniform illumination, subtracted dark counts,
and then selected a reference pixel. 
The relative detection efficiency of each pixel was determined by calculated its ratio relative to the reference.
It should be noted that the raw calibration data also include both true photon events and spurious hit pulses.

\begin{figure}[th]
\begin{minipage}{0.45\textwidth}
\includegraphics[width=0.9\textwidth]{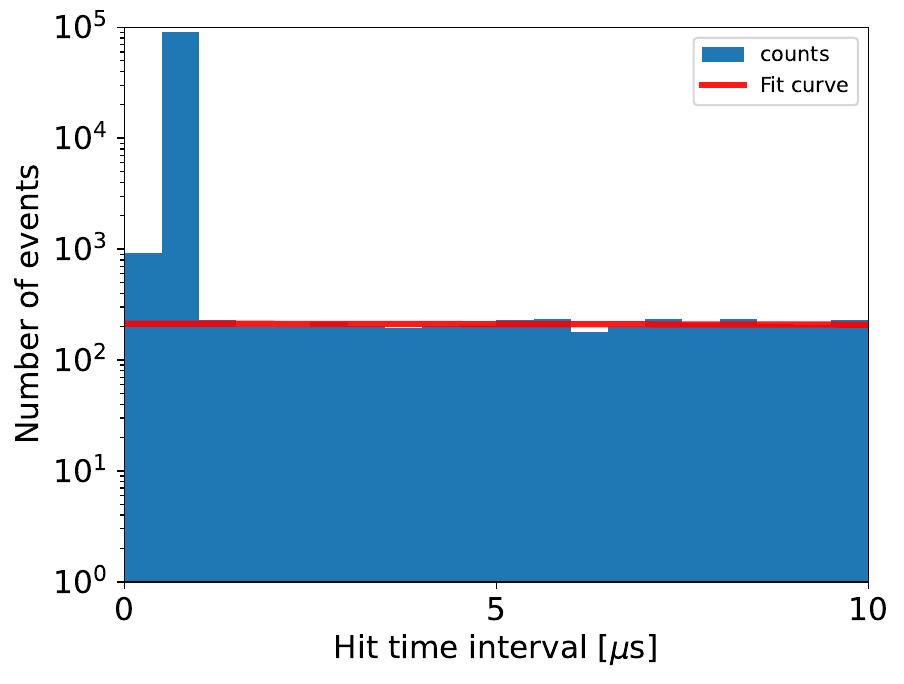}
   \end{minipage}
  \begin{minipage}{0.45\textwidth}
    \includegraphics[width=0.9\textwidth]{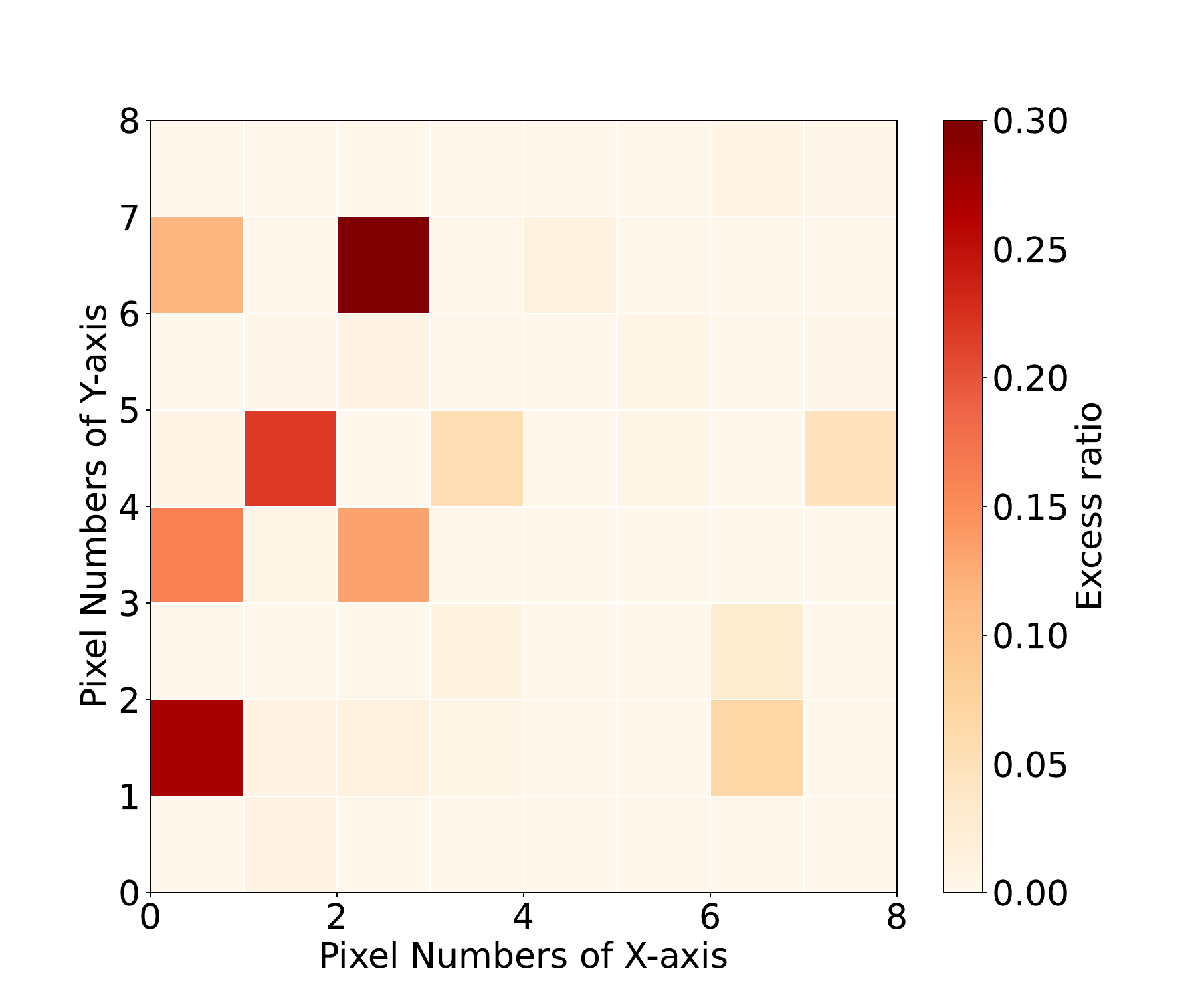}
   \end{minipage}
    \caption{(left) Close-up view of the $\Delta t$ histogram near zero,
    overlaid with the exponential fit shown as a red solid curve.
    (right) Map showing the estimated excess rate for each pixel.
        }
    \label{fig:poisson}
\end{figure}

First, we corrected this issue by estimating the number of spurious pulses $N_{\mathrm{sp}}$ based on a Poisson distribution. 
We analyzed time intervals between dark pulses in each cell. 
The average $\Delta{\,t}$ was about $100\,\mu\mathrm{s}$, 
but dropped below $1\,\mu$s in some cases due to fluctuations in the signal tail.
A hit pulse stays high while the analog signal exceeds the threshold voltage. 
If the signal tail crosses the threshold again, 
an additional short hit pulse may occur, reducing the time interval between pulses (Figure~\ref{fig:hakei}). 
We modeled the $\Delta{\,t}$ distribution with an exponential function $re^{-rt}$, 
where $r$ is the event rate from Poisson statistics. 
Figure~\ref{fig:poisson} shows our identification of spurious pulses using $\Delta{\,t}$ analysis. 
We fitted the $\Delta{\,t}$ histogram with this function (Figure~\ref{fig:poisson}, left) and identified excess events near $\Delta t \sim 0$ corresponding to $N_{\mathrm{sp}}$. 
To avoid biasing the fit, we fitted the range $1< \Delta t< 100\,\mu$s 
and extrapolated the curve to $\Delta t=0$. 
The excess above the extrapolated curve was defined as $N_{\mathrm{sp}}$.
Correction coefficients were then calculated for each cell to exclude $N_{\mathrm{sp}}$.
Figure~\ref{fig:poisson} (right) shows that only a small fraction of pixels were significantly affected. 
After correcting for $N_{\mathrm{sp}}$, we applied FFC, ensuring a more uniform response for accurate photometric measurements.

In another approach, we applied FFC without explicitly correcting for spurious hit pulses.
The flat-field data used for FFC included both true photon events and spurious contributions.
By selecting reference pixels with minimal contamination from spurious hit pulses,
FFC effectively corrected for both $N_{\mathrm{sp}}$ and intrinsic PDE variations across the array.
The essential difference between the two methods lies in whether FFC was applied after correcting spurious hit pulses or directly to the uncorrected data.
We compared the results obtained by both approaches, and 
confirmed consistency within statistical uncertainties, as expected.
This agreement validates that the effect of spurious hit pulses has been properly mitigated and confirms the robustness of the FFC-based correction strategy.




\section{Linearity}

\begin{figure}[htbp]
\centering
\begin{minipage}[b]{0.42\columnwidth}
    \centering
    \includegraphics[width=0.85\columnwidth, trim={355 105 355 105}, clip]{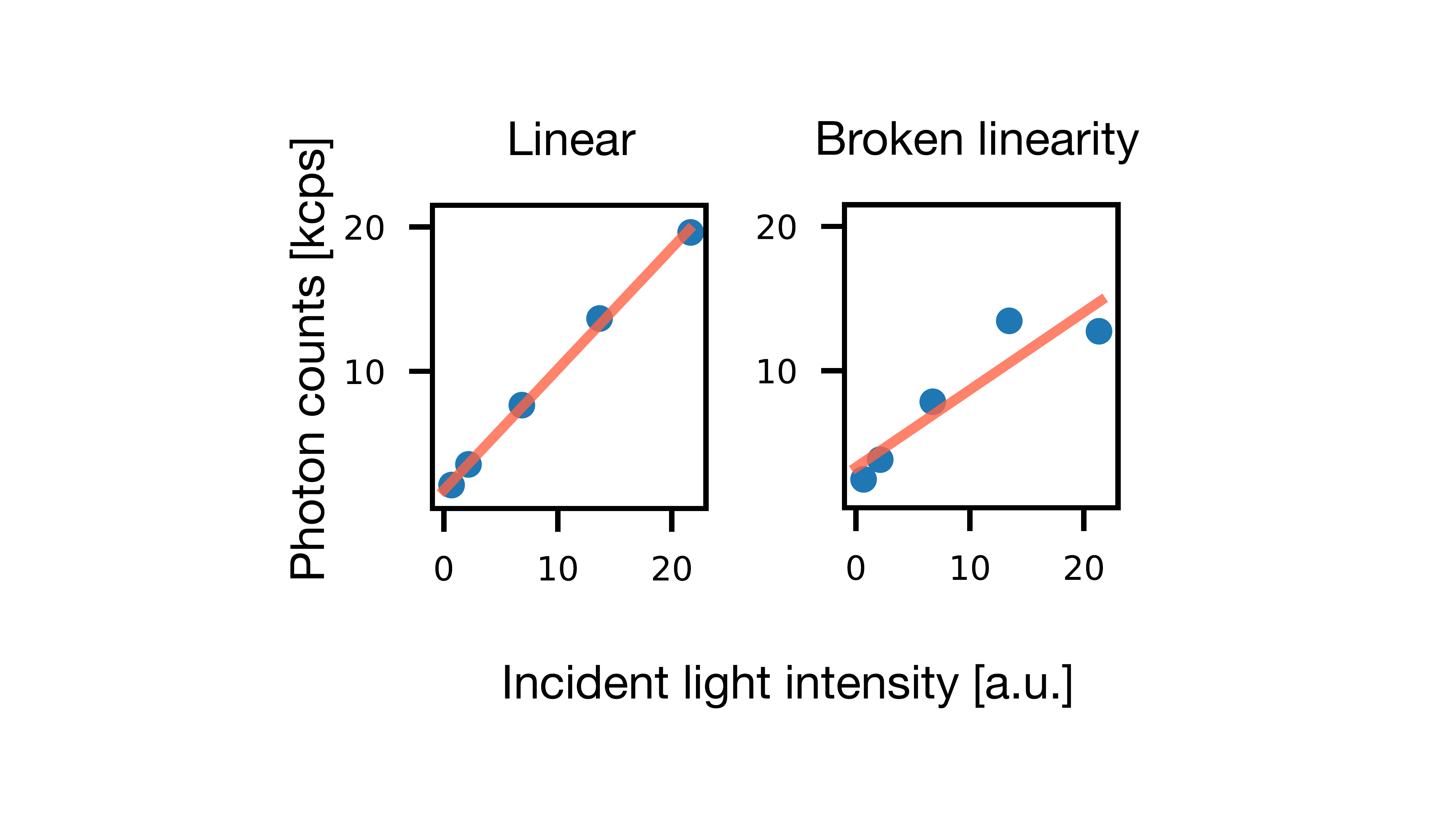}
    \caption{Examples of linearity plots for a well-behaved pixel (left) and a pixel with degraded linearity (right).}
    \label{fig:lin_gb}
\end{minipage}
\hspace{0.01\columnwidth}
\begin{minipage}[b]{0.42\columnwidth}
    \centering
    \includegraphics[width=\columnwidth]{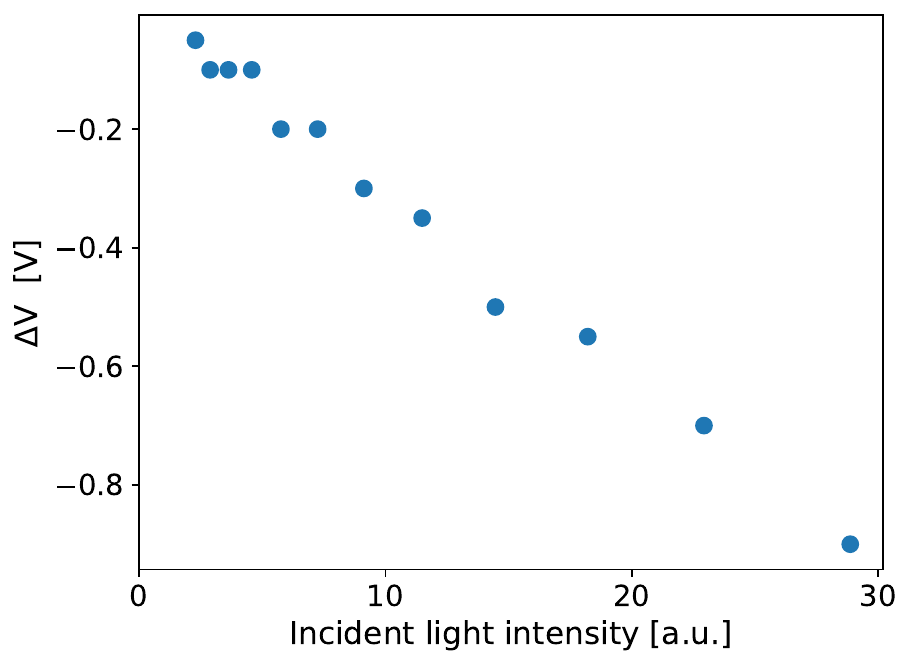}
    \caption{High-voltage decrease plotted against incident light intensity.}
    \label{fig:hv_drop}
\end{minipage}
\end{figure}

We evaluated pixel-by-pixel linearity 
by illuminating the entire sensor surface with DC light and adjusting its intensity, as shown in Figure~\ref{fig:exsetup}. 
For each of the 64 cells,
the relative incident light intensity, calculated based on the absorption coefficients of the ND filters,
was plotted on the horizontal axis,
while the number of detected photon counts was plotted on the vertical axis. 
Figure~\ref{fig:lin_gb} presents example plots for selected cells.
Most cells exhibited linear behavior,
although some deviated from linearity on the order of
10 kilo counts per second (kcps).

Further investigation revealed that exposure to intense light 
caused a drop in high voltage due to increased photocurrent.
This effect was attributed to a resistor in the low-pass filter between the high-voltage supply and the sensor.
Figure~\ref{fig:hv_drop} shows the high-voltage drop as a function of incident light intensity.
As the incident light intensity increases, the high-voltage decreases almost linearly.
The resistance value was excessively large due to insufficient optimization, leading to an unnecessary voltage drop.
In the next version, 
we plan to replace this resistor with a smaller one to prevent voltage drop and extend the linearity range.

\section{Performance on the Telescope}
\begin{figure}[htbp]
\centering
\begin{minipage}[b]{0.43\columnwidth}
    \centering
    \includegraphics[width=0.95\columnwidth]{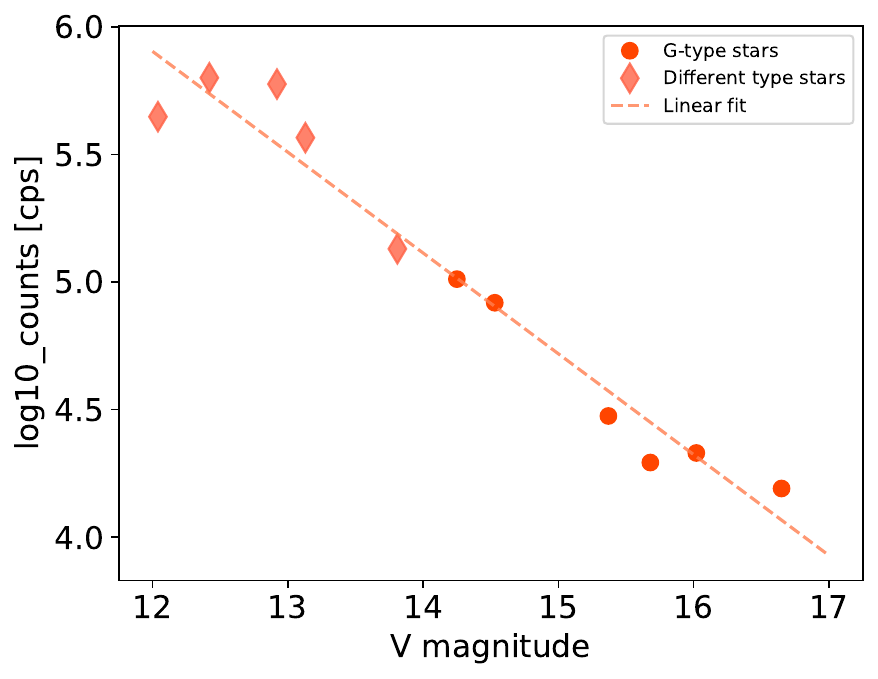}
    \caption{Detected photon counts of stars as a function of V magnitude. 
        Filled circles and diamonds indicate values for G-type stars 
        and arbitrarily selected non-G-type stars, respectively.
        The dashed line represents the linear fit to all data points.}
    \label{fig:seimei_lin}
\end{minipage}
\hspace{0.02\columnwidth}
\begin{minipage}[b]{0.43\columnwidth}
    \centering
    \includegraphics[width=0.95\columnwidth]{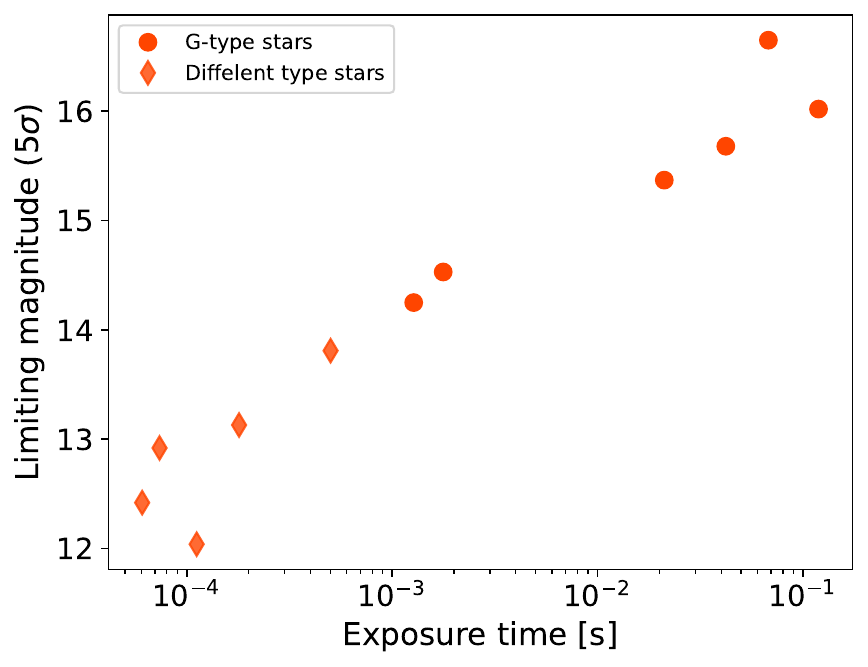}
    \caption{5$~\sigma$ limiting magnitude versus exposure time.
        Data points correspond to those in Figure\,\ref{fig:seimei_lin}.
        }
    \label{fig:sn}
\end{minipage}
\end{figure}

We observed stars with the Seimei telescope 2023 October, 2024 February, and 2024 October to evaluate the linearity with the focused light.
At first, 
we plotted the V-band magnitude on the horizontal axis 
and the number of detected photon counts on the vertical axis,
as shown in Figure~\ref{fig:seimei_lin}.
The fitted slope was $-0.395 \pm 0.029$, which is consistent within $1\sigma$
with the expected value of $-0.4$ derived from the Pogson equation,
$\Delta \log{n}/\Delta m=-0.4$, where $n$ is the photon flux and $m$ is the V magnitude.
These results suggest that differences in the number of detected photons due to the spectral type of stars have little effect.


To further assess photometric performance, 
we estimated the signal-to-noise ratio (S/N) for each stellar image.
The S/N was calculated using the standard equation
$\mathrm {S/N}=(N_{\mathrm{ON}}-N_{\mathrm{OFF}})/\sqrt{N_{\mathrm{ON}}+N_{\mathrm{OFF}}}$,
where $N_\mathrm{ON}$ is the total photon count in the signal region (16 pixels),
and $N_\mathrm{OFF}$ is the total photon count in the background region.
By slicing the data into segments of a specified time width, 
multiple images were extracted.
For each individual image the S/N was calculated, 
and the mean of its distribution provides the S/N value corresponding to that exposure time.
By varying the time width, we obtained the relationship between exposure time and S/N,
in which the data was clearly proportional to the square root of exposure time.
We then also calculated the exposure time needed to detect the star at 5$~\sigma$.
By repeating this analysis for stars of different magnitudes, 
we finally obtained a limiting magnitude plot as shown in Figure\,\ref{fig:sn}.
The primary source of the systematic uncertainty was
the variation in the sky condition and hence in the background count rate.

\section{Summary}
We reported that the optical photon-counting system IMONY worked properly with the 3.8-m Seimei telescope
especially in the context of photometry.
We evaluated the system linearity and limiting magnitude, making IMONY one of
the most advanced and unique detectors for unveiling the mysteries of short-timescale phenomena in astrophysics.
The dynamic range will be extended by optimizing the resistance in the low-pass filter in the bias circuit
in the upcoming version \cite{ref:Anju}.



\acknowledgments

This work was supported by JSPS KAKENHI Grant Numbers 23H01194.
We thank the referee for their helpful feedback, which has improved this work.





\end{document}